\documentclass[prl,english,showpacs,preprint,superscriptaddress]{revtex4}
\usepackage[english]{babel}

\usepackage{amsfonts}
\usepackage{amsmath}
\usepackage{amsthm}
\usepackage{amssymb}
\usepackage{graphicx}
\usepackage{subfigure}
\usepackage{cancel}
\usepackage[usenames,dvipsnames]{color}

\newcommand{\mm} {~\mathrm{mm}}
\newcommand{\nm} {~\mathrm{nm}}
\newcommand{\nA} {n_\mathrm{A}}
\newcommand{\nf} {n_\mathrm{f}}
\newcommand{\Kexp} {K_\mathrm{exp}}
\newcommand{\Kth} {K_\mathrm{th}}
\newcommand{\Gammae} {\Gamma_\mathrm{e}}
\newcommand{\PhiC} {\Phi_\mathrm{C}}

\usepackage{natbib}

\begin{document}

\title{Off-equilibrium surface tension in colloidal suspensions }%across the jamming transition}

\author{Domenico~Truzzolillo}
\email{domenico.truzzolillo@univ-montp2.fr}
\author{Serge~Mora}
\altaffiliation{Present address: Laboratoire de M\'{e}canique et de G\'{e}nie Civil - UMR 5508
CNRS et Universit\'{e} de Montpellier 2 Place E.Bataillon F-34095 Montpellier cedex 5}
\author{Christelle Dupas}
\author{Luca~Cipelletti}

\affiliation{Universit{\'e} Montpellier 2, Laboratoire Charles Coulomb UMR 5221,
F-34095 Montpellier, France}
\affiliation{CNRS, Laboratoire Charles Coulomb UMR 5221, F-34095 Montpellier, France}


\date{\today}

\begin{abstract}
We study the fingering instability of the interface between two miscible fluids, a colloidal suspension and its own solvent.
The temporal evolution of the interface in a Hele-Shaw cell is found to be governed by the competition between the non-linear viscosity of the suspension and an off-equilibrium, effective surface tension $\Gammae$. By studying suspensions in a wide range of volume fractions, $\Phi_\mathrm{C}$, we show that $\Gammae \sim \Phi_\mathrm{C}^2$, in agreement with Korteweg's theory for miscible fluids. The surface tension exhibits an anomalous increase with particle size, which we account for using entropy arguments.
\end{abstract}

\maketitle

The surface tension between two fluids quantifies the energetic costs of creating new interface~\cite{vanderwaals1894,Degennes}. At equilibrium, surface tension may only exist between immiscible fluids: when two \textit{miscible} fluids are brought in contact, the initial concentration gradient across the interface rapidly relaxes via diffusion and the system reaches an equilibrated, uniform state. On time scales shorter than that of interface relaxation, however, there are capillary forces at the interface that mimic an effective surface tension, as it was already recognized by Korteweg in 1901~\cite{Korteweg1901}. Similarly to the theory for immiscible fluids, Korteweg's theory relates the effective surface tension $\Gammae$ to the gradient of composition across the interface:
\begin{equation}\label{eq:sq-grad}
\Gammae=  \kappa\int_{-\infty}^{\infty} \left(\frac{d\varphi}{dz}\right)^2 dz \simeq \frac{\kappa}{\delta} \Delta\varphi^2 \,,
\end{equation}
where $z$ is the coordinate orthogonal to the interface, $\kappa$ the Korteweg constant, and $\varphi$ the concentration of one of the two species. The last approximation holds for a linear concentration profile that increases by $\Delta \varphi$ across an interface of thickness $\delta$. Subsequent work by Davis~\cite{Davis1988} and Joseph~\cite{Joseph1990} has generalized Korteweg's ideas by suggesting that interfacial stresses may arise whenever gradients of an arbitrary fluid property exist at the interface between miscible fluids, \textit{e.g.} density or temperature.

The existence of a transient effective surface tension has been demonstrated in light scattering experiments probing capillary waves at the interface between miscible fluids~\cite{May1991,cicuta01}. Korteweg stresses have also been invoked to explain the shape of drops and bubbles, both under the effect of gravity~\cite{Joseph1990} and in spinning drop measurements~\cite{petitjeans,Pojman2006,Pojman2007}, the onset of a Marangoni-like instability leading to the cellular convective mixing of miscible fluids~\cite{Jacob1991}, and the shape of the meniscus between molten silicates of different composition~\cite{Mungall1994}. In spite of the possible relevance of $\Gammae$ in many situations, including jetting, bubbles and drops formation, coalescence and break-up, plumes and convection, precipitation and deposition, experiments that quantitatively probe Korteweg's theory remain scarce: very few data are available for $\Gammae$~\cite{May1991,cicuta01,Pojman2006,Pojman2007}, and large discrepancies between experimental values and those estimated from Eq.~(\ref{eq:sq-grad}) have been reported~\cite{petitjeans}. An increase of $\Gammae$ with $\Delta \varphi$ was reported in spinning drop experiments on water-glycerin~\cite{petitjeans} and polymer~\cite{Pojman2007} systems, but large deviations with respect to the quadratic scaling of Eq.~(\ref{eq:sq-grad}) were observed. The very existence of an off-equilibrium surface tension is debated. Numerical simulations~\cite{Chen2008} of the fingering instability arising when a less viscous fluid is pushed through a miscible, more viscous one in a Hele-Shaw cell highlight the role of $\Gammae$ in stabilizing the interface. By contrast, in earlier works the observed patterns were explained without including the contribution of the surface tension~\cite{Paterson1985}, or by explicitly assuming $\Gammae = 0$~\cite{Nittman1985,Nittman1986}.

Colloidal suspensions may be regarded as ideal benchmark systems to investigate surface tension effects, thanks to the possibility of controlling the interparticle interactions and because the interface may be probed in great detail, down to the particle level~\cite{Aarts2004}. Previous work has focussed on the equilibrium interface between phase-separated colloidal fluids~\cite{dehoog99,Aarts2004,Lekkerkerker08,aarts13}; however, colloidal suspensions are also excellent candidates for investigating off-equilibrium surface tension. Indeed, diffusion is much slower in colloids as compared to atomic systems, leaving a wider temporal window for probing the transient interface between miscible fluids. Additionally, the rich rheological behavior of colloidal suspensions allows one to explore surface tension beyond the simple case of Newtonian fluids typically relevant for molecular fluids.

In this letter, we report Hele-Shaw experiments on the fingering instability observed at the interface between two miscible fluids, a colloidal suspension and its own solvent. We show that the evolution of the interface pattern is governed by both the non-linear viscosity of the suspension and an effective surface tension, which we measure as a function of the volume fraction of the suspension. Our results confirm the quadratic scaling predicted by Korteweg, Eq.~(\ref{eq:sq-grad}). We furthermore show that, for our microgel particles, $\Gammae$ is governed by the entropy associated with the internal degrees of freedom of the particles, leading to a surprising, previously unreported growth of $\Gammae$ with particle size.

The experiments are performed in a Hele-Shaw cell consisting of two square glass plates of side $L=25\mm$ separated by four Mylar spacers, fixing the gap at $b=0.5\mm$. The cell is filled with the fluid to be studied, whose viscosity is $\eta_2$. A less viscous fluid is injected through a hole of radius $r_0=0.5\mm$ in the center of the top plate. For all experiments, we use water died with 0.5\% w/w of methylene blue as the less viscous fluid, with viscosity $\eta_1 = 1.011~\mathrm{mPa~s}$. The injected volume per unit time, $\dot{V}$, is controlled via a syringe pump. Temperature is fixed at $T = 293\pm 0.1~\mathrm{K}$ by means of a Peltier element placed under the bottom glass plate. The Peltier has a circular hole of radius $8.5\mm$ for optical observation. A fast CMOS camera (Phantom v7.3 by Vision Research) run at 100 to 3000 frames $\mathrm{s}^{-1}$ is used to record movies during injection, by imaging the sample through the bottom plate.

\begin{figure}[htbp]
\centering
\includegraphics[width=0.9\columnwidth,clip]{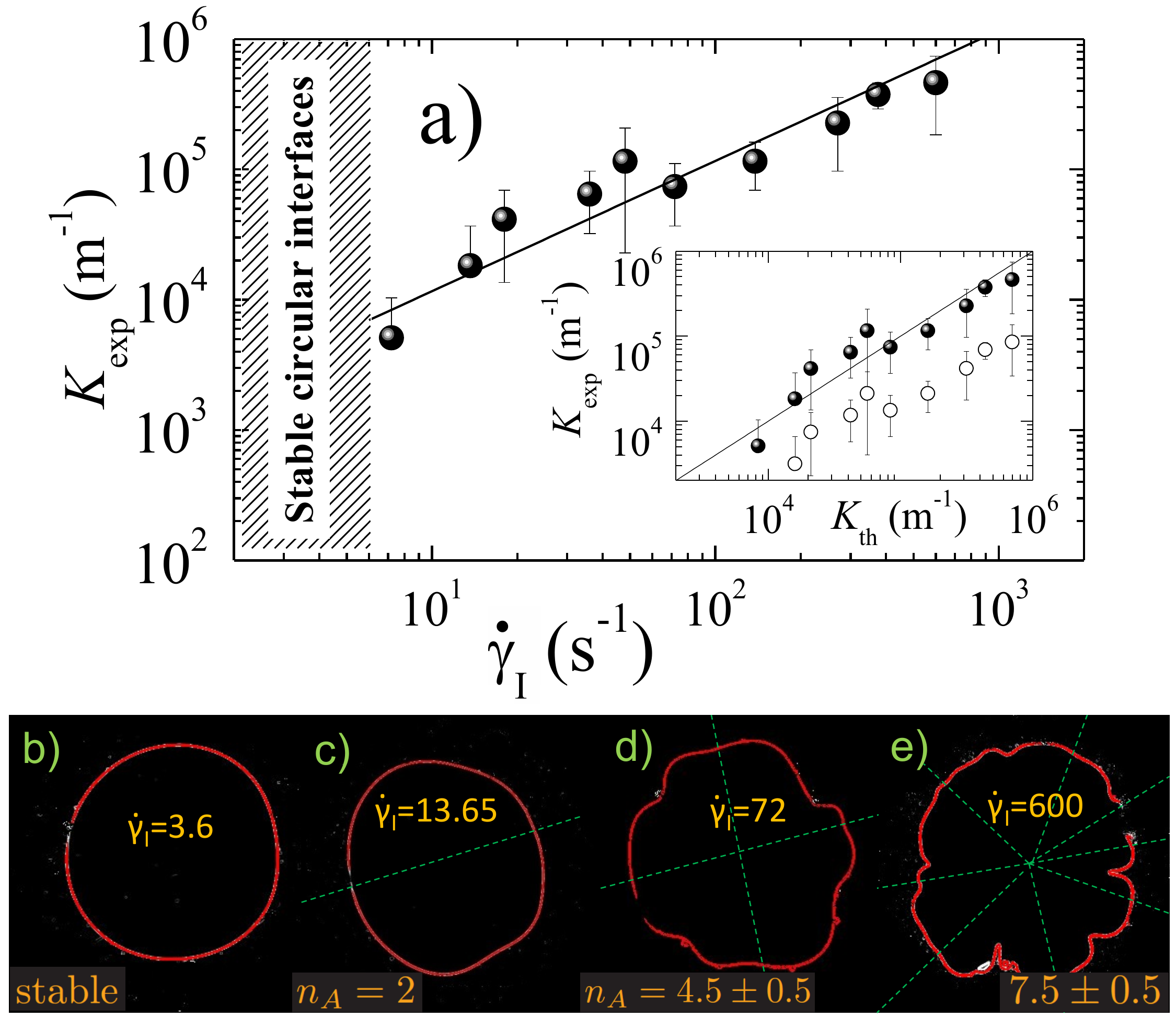}\\
\caption{(Color online) a): Finger function $\Kexp$ \textit{vs} $\dot{\gamma}_\mathrm{I}$ in Hele-Shaw experiments where water is injected in silicon oil. The solid line is the theoretical function $\Kth$ (r.h.s. of Eq.~(\ref{eq:Kappafun3})), with no adjustable parameters. Inset: $\Kexp$ \textit{vs} $\Kth$, as obtained from the analysis proposed in the text (solid circles) or using previous approaches based on the fastest-growing Fourier mode of the instability (open circles). b)-e) Water-oil interface (red line), as observed right after the onset of fingering, for various injection shear rates (in $\mathrm{s}^{-1}$) as indicated by the labels.}
\label{fig:Silicon-water}
\end{figure}

\begin{figure}[htbp]
\centering
\includegraphics[width=0.9\columnwidth,clip]{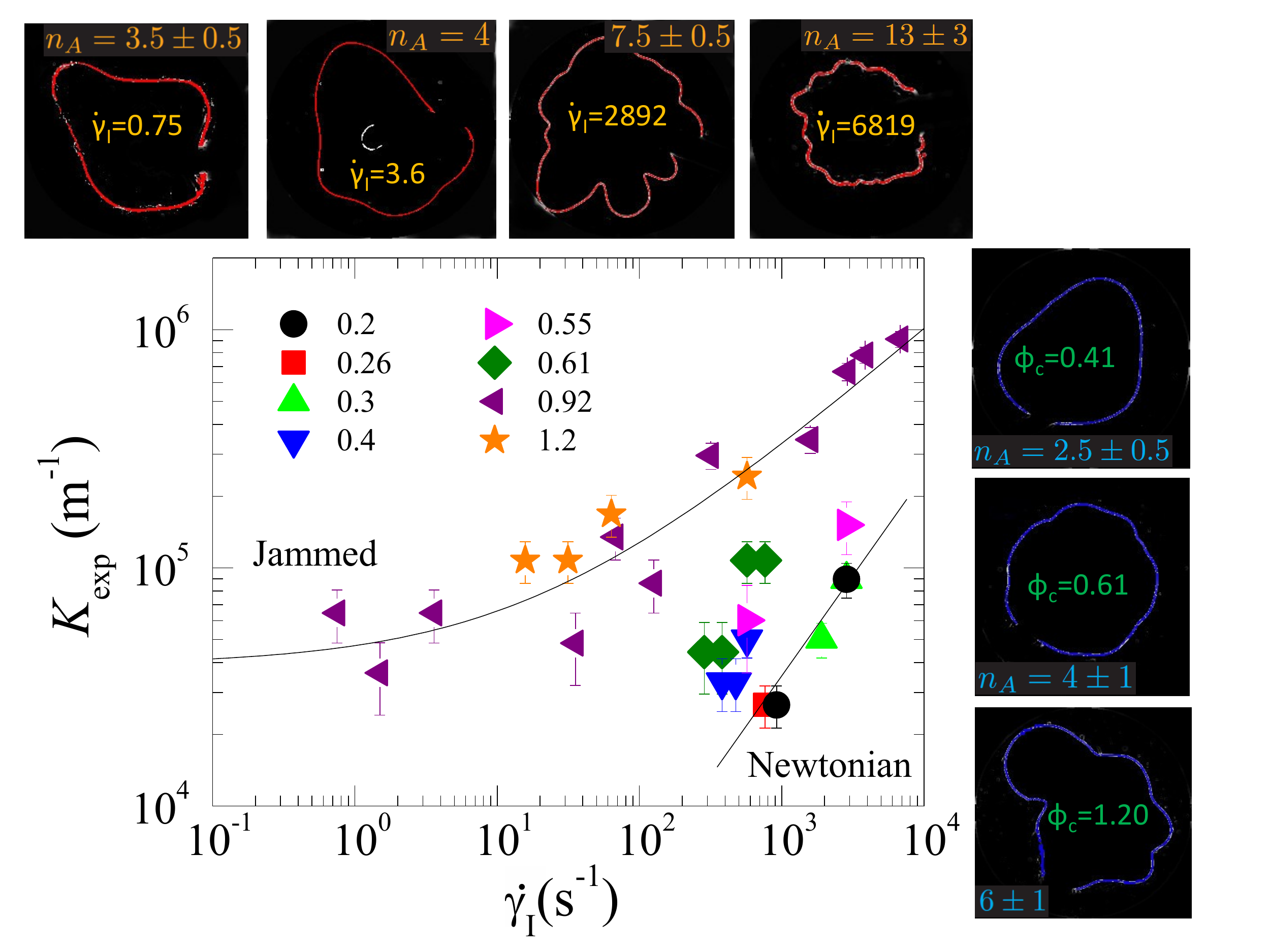}\\
\caption{(Color online). Finger function $\Kexp$ \textit{vs} $\dot{\gamma}_\mathrm{I}$ measured when injecting water in a microgel suspension. The labels indicate the suspension volume fraction. The lines are guides for the eye. Small panels: water-suspension interfaces for $\Phi_\mathrm{C}=0.92$ and various $\dot{\gamma}_\mathrm{I}$ (top row, as indicated by the labels, in $s^{-1}$), or at fixed $\dot{\gamma}_\mathrm{I} = 570~\mathrm{s}^{-1}$ and various $\Phi_\mathrm{C}$ (right column).}
\label{fig:K-pnipam}
\end{figure}

Typical images of the interface between the two fluids are shown in Figs.~\ref{fig:Silicon-water} and~\ref{fig:K-pnipam}, where the distinctive instabilities that develop when $\eta_1<\eta_2$ are clearly visible. In the framework of linear evolution theory, such an instability is conveniently described by decomposing the interface profile in Fourier modes, the mode of order $n$ being associated with a pattern with $n$ lobes, or fingers. For two Newtonian fluids, the order $n_\mathrm{f}$ of the mode with the fastest growth rate is given by~\cite{Miranda1998}
\begin{equation}\label{eq:fastmode}
n_\mathrm{f}=\frac{1}{\sqrt{3}}\left[\frac{4r r_0  \dot{\gamma}_\mathrm{I}(\eta_2-\eta_1)}{b \Gamma}+1\right]^{0.5} \,,
\end{equation}
where $r=\left[r_0^2+\dot{V} t/(\pi b)\right]^{0.5}$ is the time-dependent radius of the unperturbed interface, $\Gamma$ the interfacial tension between the two fluids and $\dot{\gamma}_\mathrm{I}=3\dot{V}(2\pi r_0 b^2)^{-1}$ the shear rate at the injection hole.
This expression is often used to describe the number of fingers experimentally observed at the onset of the instability~\cite{Alvarez2004}. However, in experiments the observed number of fingers should be regarded as being related to the mode with maximum amplitude, rather than to the fastest growing one. We thus modify the standard linear evolution theory to calculate $n_\mathrm{A}$, the order of the experimentally accessible mode with the maximum amplitude, finding~\cite{note}
\begin{equation}\label{eq:maxampl2}
n_\mathrm{A}=\alpha n_\mathrm{f} = \frac{\alpha}{\sqrt{3}}\left[\frac{4r\dot{\gamma}_\mathrm{I}(\eta_2-\eta_1)}{r_0 b \Gamma}+1\right]^{0.5}\,,
\end{equation}
where $\alpha=\sqrt{-W\left(-3e^{-3}\right)}\simeq 0.422$, with $W(x)$ the Lambert function satisfying $x=W(x)e^{W(x)}$. Although Eq.~(\ref{eq:maxampl2}) is formally derived in the limit $n_\mathrm{A}>>1$, we check numerically that it holds to a very good approximation already for $n_\mathrm{A}\ge 2$~\cite{note}.

We test the validity of Eq.~(\ref{eq:maxampl2}) by performing Hele-Shaw experiments using two Newtonian fluids for which all the relevant parameters are known: dyed water and silicon oil ($\eta_2 = 12.5~\mathrm{Pa~s}$, $\Gamma = 39.8~\mathrm{mN~m}^{-1}$~\cite{Koos2012}). Figures~\ref{fig:Silicon-water}(b-e) show typical interface patterns observed at various injection rates. We determine $\nA$ by counting the number of fingers of the destabilized interface, averaging over typically two or three independent experiments for each $\dot{\gamma}_\mathrm{I}$. To compare the experiments to the theory, it is convenient to recast Eq.~(\ref{eq:maxampl2}) in the form
\begin{equation}\label{eq:Kappafun3}
\Kexp(\nA,r)= \Kth \,,
\end{equation}
where the experimental `finger function' $\Kexp$ is defined by
\begin{equation}\label{Kappafun1}
\Kexp(\nA,r)=\frac{1}{r}\left[\frac{3\nA^2}{\alpha^2}-1\right] \,,
\end{equation}
while its theoretical value depends only on the rheological and interfacial properties of the fluids, the cell geometry, and the imposed shear rate:
\begin{equation}\label{eq:Kappatheo}
\Kth = \dot{\gamma}_\mathrm{I}\frac{4r_0(\eta_2-\eta_1)}{b\Gamma}\,.
\end{equation}

Figure~\ref{fig:Silicon-water}(a) shows $\Kexp(\nA,r)$ \textit{vs} $\dot{\gamma}_\mathrm{I}$ for the water-silicon oil system. The experimental points (symbols) are in excellent agreement with the line, which shows $\Kth$, obtained from Eq.~(\ref{eq:Kappatheo}) using the fluids and cell parameters. We emphasize that such a quantitative agreement would not hold if $\Kexp$ was calculated by interpreting the observed number of fingers as the fastest-growing Fourier mode of the destabilized interface, \textit{i.e.} if $\nA$ was replaced by $\nf = \nA/\alpha$ in Eqs.~(\ref{eq:Kappafun3},\ref{Kappafun1}), as shown in the inset of Fig.~\ref{fig:Silicon-water}(a).
%, where $\Kexp$ calculated using $\nA$ (solid symbols) or $\nf$ (open circles) is plotted against $\Kth$. Clearly, a significant discrepancy is observed when using $\nf$.

Having demonstrated that our experiments allow the flow and interfacial parameters to be quantitatively determined, we use the same setup to investigate the off-equilibrium interfacial tension between a colloidal suspension and its solvent. We study aqueous suspensions of poly-N-isopropylacrylamide (PNiPAM) microgel particles, whose synthesis is described in Ref.~\cite{Senff1999}. At $T=293~\mathrm{K}$ the particles have hydrodynamic radius $R_\mathrm{h}=165\nm$, as measured by dynamic light scattering in dilute suspensions. For the same synthesis, the radius of gyration has been determined to be $R_\mathrm{g}\simeq 0.5 R_\mathrm{h}$~\cite{Sessoms2009}. We perform experiments for several particle concentrations, which we express as the effective volume fraction, $\Phi_\mathrm{C}$, of the microgels. Experimentally, the polymer mass concentration, $c$ (w/w), is known from the synthesis. In the dilute regime where direct particle interactions are negligible, the effective volume fraction is simply proportional to $c$. For all concentrations, we define the effective volume fraction as $\Phi_\mathrm{C} = k  c$, where $k$ is determined from the viscosity of the suspension in the dilute limit. We find $k = 20.1$ by matching the $c$-dependent zero-shear viscosity of the suspension to Einstein's formula, $\eta = \eta_0 (1 + 2.5 kc)$, where $\eta$ is the viscosity of the suspension and $\eta_0$ that of the solvent. Viscosity measurements are performed using an Anton Paar Lovis 2000 ME microviscosimeter, in the range $0 < \Phi_\mathrm{C} \le 0.02$.
Our Hele-Shaw experiments cover suspensions with volume fractions ranging from $\Phi_\mathrm{C}= 0.2$, corresponding to diluted, hard sphere-like suspensions, up to $\Phi_\mathrm{C}  = 1.2$, where particles are squeezed due to steric constraints and the suspension is fully jammed. For a given $\Phi_\mathrm{C}$, we perform experiments at various $\dot{\gamma}_\mathrm{I}$, always keeping the injection rate high enough for diffusion-driven mixing between the injected solvent and the suspension to be negligible~\cite{noteinjection}.

Figure~\ref{fig:K-pnipam} shows $\Kexp$ as a function of the injection shear rate for all the microgel suspensions. A qualitative change is observed when $\Phi_\mathrm{C}$ increases: at low volume fraction $\Kexp~\sim \dot{\gamma}_\mathrm{I}$, while for jammed suspensions $\Kexp$ grows sublinearly with $\dot{\gamma}_\mathrm{I}$ at high shear rate and tends to a plateau for $\dot{\gamma}_\mathrm{I} \rightarrow 0$. This behavior is strongly reminiscent of the shape of the flow curve, $\sigma(\dot{\gamma})$, in colloidal suspensions, where $\sigma = \eta \dot{\gamma}$ is the shear stress when imposing a shear rate $\dot{\gamma}$. This suggests that $\Kexp$ is proportional to the shear stress contrast, \textit{i.e.} that Eq.~(\ref{eq:Kappafun3}) may be generalized by
\begin{equation}\label{eq:Kappa-nonNewton}
\Kexp= \frac{4r_0\dot{\gamma}_\mathrm{I}}{b}\frac{\eta_2(\dot{\gamma}_r)-\eta_1}{\Gammae}\,,
\end{equation}
where $\eta_2(\dot{\gamma}_r)$ is the shear-rate dependent viscosity of the suspension, $\dot{\gamma}_r=\frac{4r_0\dot{\gamma}_I}{r}$ the shear rate at the position $r$ of the interface (assuming Poiseuille flow), and $\Gammae$ the ($\Phi_\mathrm{C}$-dependent) effective surface tension between the suspension and its solvent. In writing Eq.~(\ref{eq:Kappa-nonNewton}) one implicitly assumes that the same kind of patterns are observed for our shear-thinning concentrated microgel suspensions as for Newtonian fluids. Numerical work on the Saffman-Taylor instability in a radial Hele-Shaw geometry supports this scenario~\cite{Sader1994}, by showing that the non-Newtonian character of the fluids does not change qualitatively the instability, but just accelerates (resp., delays) its onset for shear-thinning (resp., shear-thickening) fluids. The choice of Eq.~(\ref{eq:Kappa-nonNewton}) is also supported by previous works~\cite{Bonn2005,Lindner2000} on the Hele-Shaw instability between immiscible non-Newtonian fluids in a rectangular geometry, where the dynamics of the fingers was described by a generalized Darcy law where the Newtonian viscosity was replaced by the shear rate-dependent viscosity.

\begin{figure}[htbp]
\centering
\includegraphics[width=0.9\columnwidth,clip]{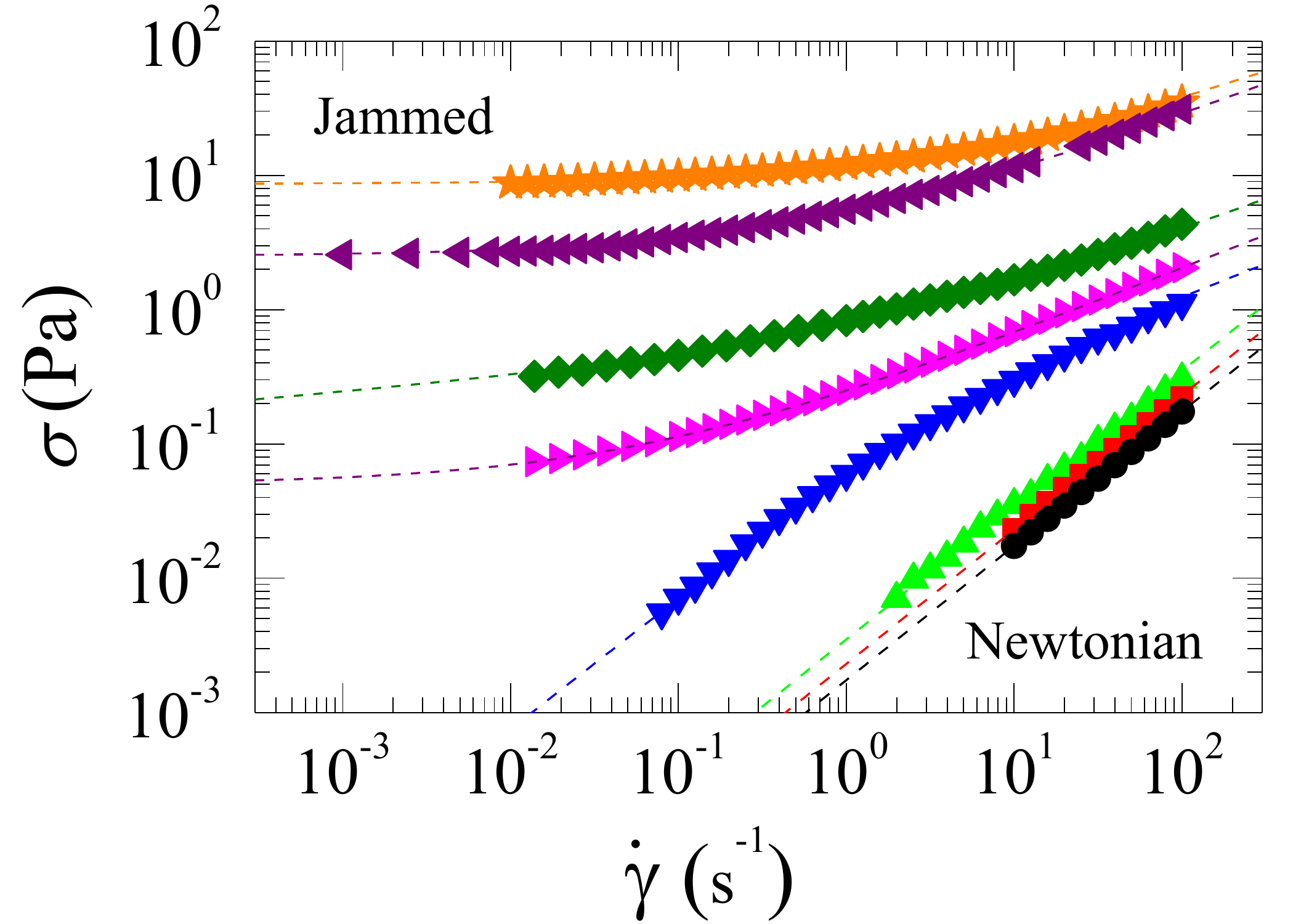}\\
\caption{(Color online). Flow curves for the microgel suspensions used in the Hele-Shaw experiments (same symbols as in Fig.~\ref{fig:K-pnipam}). The lines are fits as described in~\cite{note}.}
\label{fig:Flow}
\end{figure}

\begin{figure}[htbp]
\centering
\includegraphics[width=0.9\columnwidth,clip]{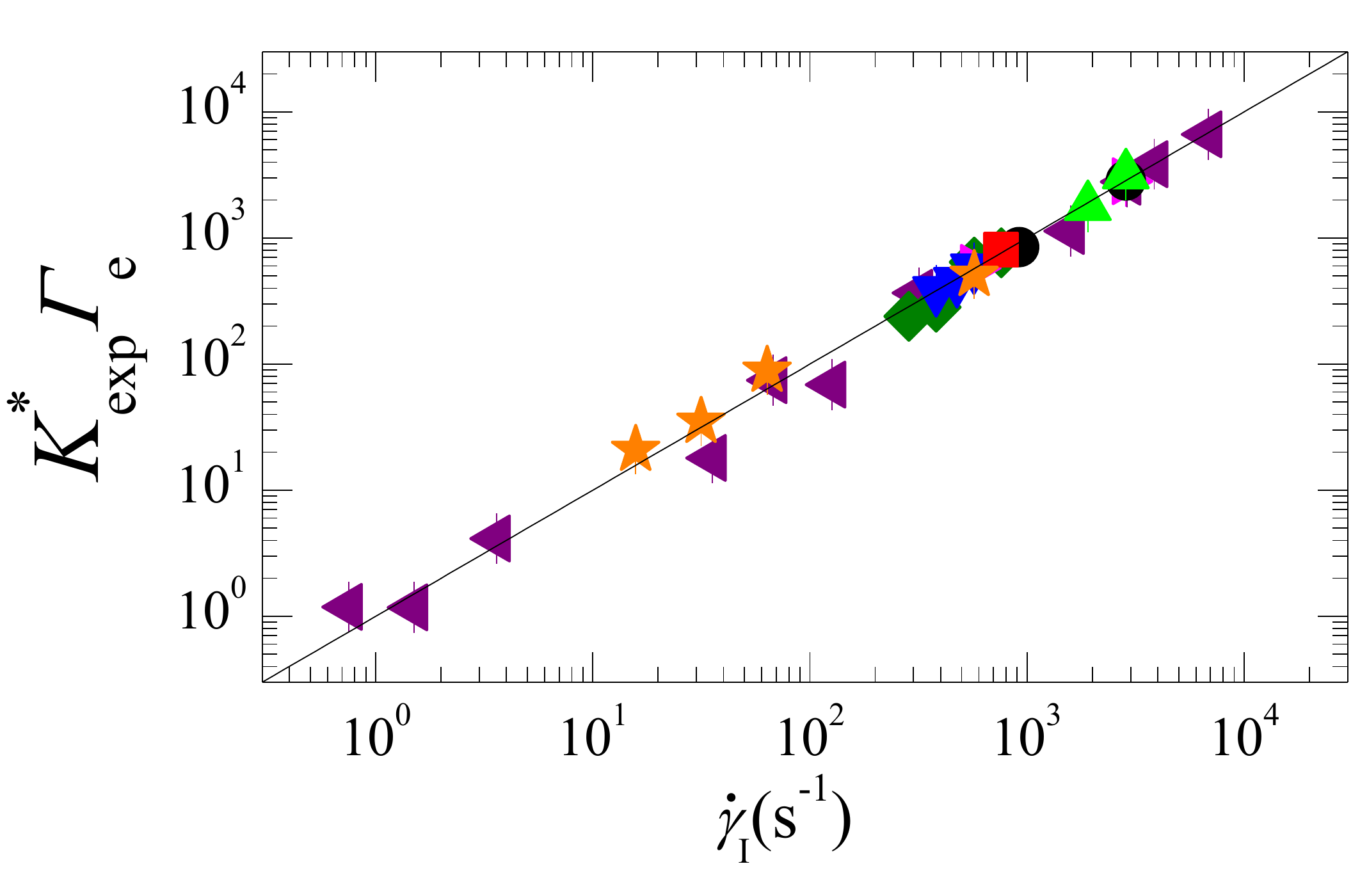}\\
\caption{(Color online). Scaled finger function $\Kexp \Gammae b\left[4r_0(\eta_2(\dot{\gamma}_r)-\eta_1)\right]^{-1}$ for the microgel suspensions whose raw data are shown in Fig.~\ref{fig:K-pnipam}. All data collapse on $\Kexp^* \Gammae =\dot{\gamma}_\mathrm{I}$ (line), thus confirming Eq.~(\ref{eq:Kappa-nonNewton}).}
\label{fig:scaling}
\end{figure}

In order to test Eq.~(\ref{eq:Kappa-nonNewton}), we measure the flow curves of the microgel suspensions. Figure~\ref{fig:Flow} shows $\sigma(\dot\gamma)$ obtained via conventional rheology. The required shear-dependent viscosity is obtained from $\eta_2 = \sigma_\mathrm{fit} \dot{\gamma}^{-1}$, where $\sigma_\mathrm{fit}$ is a fit to the measured flow curve (lines in Fig.~\ref{fig:Flow}). The fits allow the viscosity to be estimated by extrapolation in the whole range of shear rates relevant to the Hele-Shaw experiments, beyond those accessible by rheology. Standard models for Newtonian and non-Newtonian fluids are used for the fits, according to the volume fraction of the suspension: Herschel-Bulkley~\cite{Cloitre2003}, double power law~\cite{Erwin2010}, Cross-like law~\cite{Sessoms2009}) and Newtonian behavior, as detailed in Ref.~\cite{note}. We find that for all $\Phi_\mathrm{C}$ the reduced finger function, $\Kexp^* \equiv \Kexp b\left[4r_0(\eta_2(\dot{\gamma}_r)-\eta_1)\right]^{-1}$, is proportional to $\dot{\gamma}_\mathrm{I}$, as predicted by Eq.~(\ref{eq:Kappa-nonNewton}), and we determine the
proportionality coefficient $\Gammae^{-1}$ by linear fitting. Figure~\ref{fig:scaling} shows $\Kexp^* \Gammae$ as a function of the injection shear rate. When using this reduced variable, the data for all the investigated volume fractions previously shown in Fig.~\ref{fig:K-pnipam} fall onto a straight line spanning more than three orders of magnitude, thereby validating Eq.~(\ref{eq:Kappa-nonNewton}).

\begin{figure}[htbp]
\centering
\includegraphics[width=0.9\columnwidth,clip]{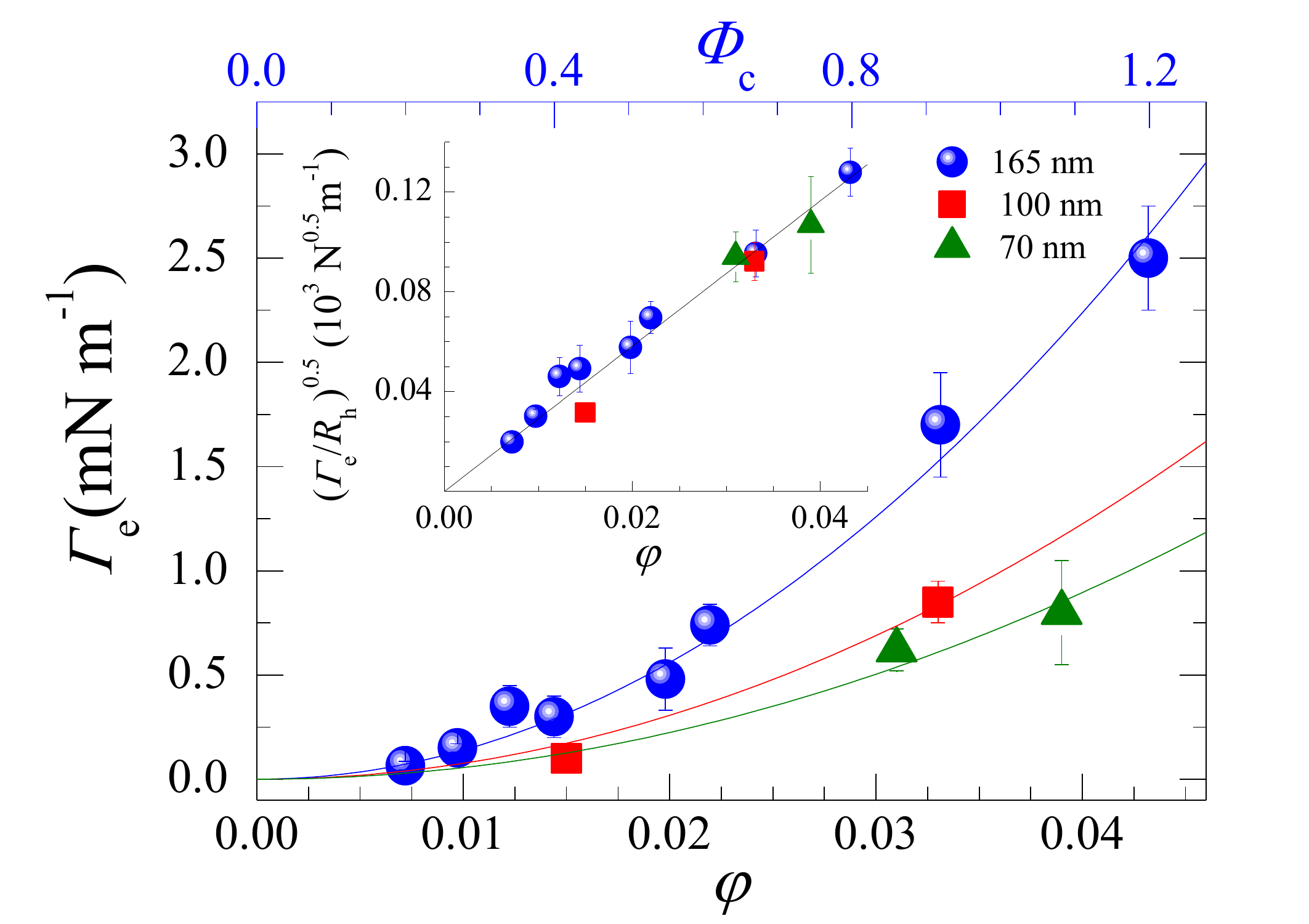}\\
\caption{(Color online). Main plot: effective interfacial tension $\Gammae$ between the microgel suspensions and their solvent, as a function of colloid (resp., polymer) volume fraction (top, resp. bottom, axis). The lines are quadratic fits to the data for microgels with various $R_\mathrm{h}$, as shown by the legend. The conversion factor between $\Phi_\mathrm{c}$ and $\varphi$ varies slightly with $R_\mathrm{h}$: the scale on the $\Phi_\mathrm{c}$ axis is exact only for $R_\mathrm{h}=165~\mathrm{nm}$. Inset: square root of the reduced surface tension $\Gammae/R_\mathrm{h}$ \textit{vs} $\varphi$, showing the collapse of all data onto a single straight line.}
\label{fig:gamma-phip}
\end{figure}

We test Korteweg's prediction, Eq.~(\ref{eq:sq-grad}), in Fig.~\ref{fig:gamma-phip}. For our system, the particle volume fraction in the injected phase is zero, so that Eq.~(\ref{eq:sq-grad}) reduces to $\Gammae \sim \Phi_\mathrm{C}^2$, where we have assumed a linear variation of the concentration profile across an interface of thickness $\delta$. Figure~\ref{fig:gamma-phip} shows that Korteweg's law holds over a wide range of concentrations, corresponding to a variation of $\Gammae$ of more than one decade. We go a step further and model our experiments at a microscopic level by calculating $\kappa$. To this end, we identify $\varphi$ in Eq.~(\ref{eq:sq-grad}) with the volume fraction of the polymer, rather than that of the particles, since the microgels are highly swollen by the solvent. Using literature values for the polymer mass density, one has $\varphi = 3.6 \times 10^{-2} \Phi_\mathrm{C}$. The Korteweg constant has been calculated by Balsara and Nauman~\cite{Balsara1988} for inhomogeneous mixtures of a solvent and ideal-chain polymers. We extend their calculation to crosslinked polymers~\cite{note}, in the limit $\varphi << 1$ relevant to our microgels, finding
\begin{equation}\label{eq:Balsarakappa}
\kappa=\frac{RTR_g^2}{6V_\mathrm{w}}\left[ \chi+3 \right]\,,
\end{equation}
with $R$ the gas constant, $V_\mathrm{w}=18 \times 10^{-6}~\mathrm{m}^3~\mathrm{mol}^{-1}$ the water molar volume and $\chi$ the Flory-Huggins parameter. Reported values of $\chi$ for PNIPAM microgels in water range from 0.25 to 0.5~\cite{Weitz2011,Hino1988,Wu2003}, yielding $5.1 \times 10^{-7} ~\mathrm{N} \leq \kappa \leq 5.49 \times 10^{-7}~\mathrm{N}$. By fitting the experimental $\Gammae$ \textit{vs} $\varphi$ we get $\kappa/\delta=1.40~\mathrm{Nm^{-1}}$ and hence $364~\mathrm{nm} \leq \delta \leq 392~\mathrm{nm}$. The interface thickness thus calculated is in very good agreement with the average distance between particles, which in the range of $\Phi_\mathrm{C}$ studied here varies from 340 nm to 460 nm. This confirms that in our experiments diffusion at the interface is negligible and validates quantitatively our analysis. To test the robustness of Eqs.~(\ref{eq:sq-grad}) and (\ref{eq:Balsarakappa}), we perform additional experiments on microgels with the same composition but smaller size, $R_\mathrm{h} = 70~\mathrm{nm}$ and $100~\mathrm{nm}$, respectively. From Eqs.~(\ref{eq:sq-grad}) and (\ref{eq:Balsarakappa}) and using $\delta \sim R_\mathrm{g} \sim R_\mathrm{h}$, one expects that data for microgels with different $R_\mathrm{h}$ should collapse onto a mastercurve when normalizing $\Gammae$ by $R_\mathrm{h}$. The inset of Fig.~\ref{fig:gamma-phip} shows that this is indeed the case. We emphasize that the scaling $\Gammae \sim R_\mathrm{h}$ is in stark contrast with the usual scaling of the interfacial tension between molecular or colloidal phases, where $\Gamma \sim a^{-2}$, with $a$ the particle size~\cite{Degennes,dehoog99}. This highlights the different origin of the surface tension in our experiments, where the entropic contribution due to the internal degrees of freedom of the polymeric particles dominates, as opposed to molecular materials where the surface tension is proportional to the particle bond energy per unit area, leading to $\Gamma \sim a^{-2}$, or the hard sphere systems of Refs.~\cite{dehoog99,Aarts2004,Lekkerkerker08}, for which translational entropy dominates, yielding $\Gamma \approx k_\mathrm{B} T/a^{2}$~\cite{noteGamma}.

In conclusion, we have investigated the pattern formation resulting from the injection of the solvent in a colloidal suspension, a model system for investigating the non-equilibrium, effective surface tension between miscible fluids. The observed interface instability can be rationalized by a remarkably simple expression, which depends separately on the rheological properties of the suspension and on the effective, off-equilibrium suspension-solvent surface tension. Our results confirm Korteweg's law and raise challenging questions on the behavior of $\kappa$, and thus $\Gammae$, as a function of inter-particle and particle-solvent interactions, as well as particle size and shape. More generally, our findings provide an experimental and theoretical framework for exploring non-equilibrium surface tension effects, a topic relevant in many problems, ranging from material processing to fundamental fluid dynamics.

\begin{acknowledgments}{This work has been supported by ANR under contract No. ANR-2010-BLAN-0402-1. The authors are grateful to E. Bouchaud, O. Dauchot, C. Ligoure, L. Ramos, and V. Trappe for useful discussions.}
\end{acknowledgments}

\newpage

\section{Supplementary Material}

We provide here details on \textit{i)} the calculation of the mode $n_A$ with maximum amplitude in the Saffman-Taylor instability in a radial Hele-Shaw geometry; \textit{ii)} the fits to the flow curves, Fig. 3 of the main text; \textit{iii)} the determination of the square gradient (Korteweg) constant $\kappa$ for microgel particles composed of cross-linked polymers.

\subsection{Saffman-Taylor instability: mode with the maximum amplitude in a radial Hele-Shaw geometry}

We start from the linear analysis of the Saffman-Taylor instability in radial geometry performed by Miranda and Widom~\cite{Miranda1998}, where the perturbation around a circular interface due to the instability is decomposed in Fourier modes of (complex) amplitude $\zeta_n(t)$. Assuming that the noise giving rise to the instability is a complex number $\zeta_n^0$, with a random phase and a $n$-independent modulus, the time-dependent amplitude of the $n$-th mode of the perturbation can be written as:
\begin{equation}\label{zeta}
%\tag{SM1}
\zeta_n(t)=\zeta_n^0\left\{\left(K(t)\frac{(nA-1)}{n^2(n-1)}^{nA-1}\right)\exp\left[(nA-1)\left(\frac{1}{K(t)}\frac{n(n^2-1)}{nA-1}-1\right)\right]\right\} \,.
\end{equation}
In Eq.~(\ref{zeta}), $A=(\eta_2-\eta_1)/(\eta_2+\eta_1)>0$ is the viscosity contrast between the two fluids, and $K(t)=[r(t)Q]/(2\pi\beta)$, where $r(t)$ is the distance from the center of the cell of the unperturbed fluid-fluid interface, $Q$ is the area covered by the injected fluid per unit time, and $\beta=b^2\Gamma/[12(\eta_1+\eta_2)]$, with $b$ the cell gap and $\Gamma$ the interfacial tension between the two fluids. Note that Eq.~(\ref{zeta}) only holds for $nA >1$. In our experiments this is not a limiting condition, since A is such that this inequality is fulfilled for $n\ge 1$ for the water-silicon oil system and for $n > 1$ for the water-microgel system.

\begin{figure}[htbp]
\centering{
\includegraphics[width=0.5\columnwidth,clip]{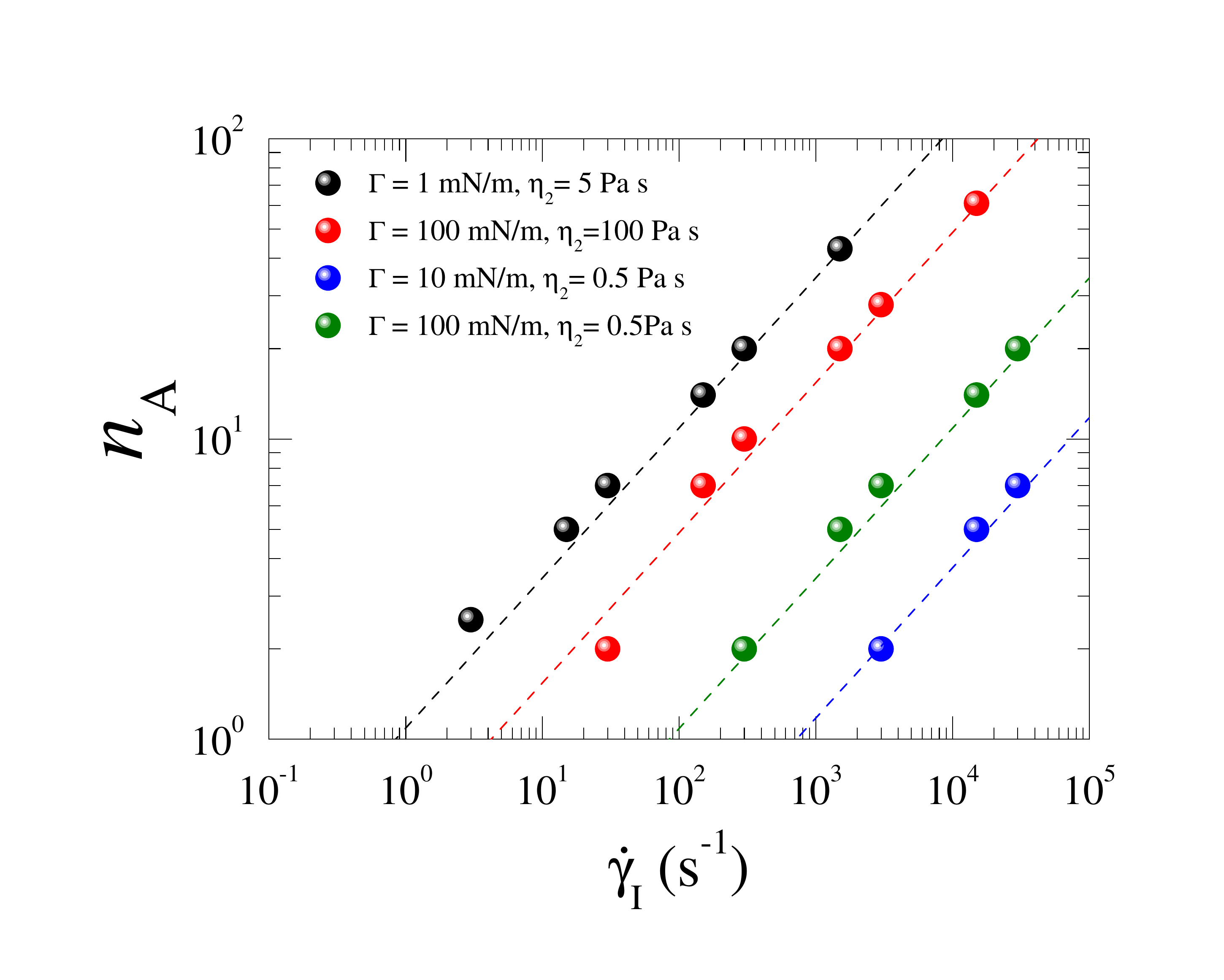}\\}
\textbf{Supplementary Figure S1:} Comparison between the exact solution for the mode with maximum amplitude (symbols) and the asymptotic approximation (dashed lines, Eqs.~(\ref{eq:nf}) and (\ref{nA})), for different values of $\eta_2$ and $\Gamma$ as shown in the legend. For all data, $\eta_1=1$~mPa~s, $b=0.5$ mm, $r_0\equiv r(0)=1$ mm. The mode number is plotted as a function of $\dot{\gamma}_\mathrm{I}=3\dot{V}(2\pi r_0 b^2)^{-1}$, the shear rate at the injection hole as defined in the main text.
\label{microgel}
\end{figure}

We calculate the mode having the maximum amplitude at a distance $r$ from the center of the Hele-Shaw cell by solving
\begin{equation}\label{der0}
%\tag{SM2}
\frac{d\zeta_n(t)}{dn}=0\,.
\end{equation}
By substituting Eq.~(\ref{zeta}) in Eq.~(\ref{der0}), one finds that the number of the mode with maximum amplitude must satisfy \begin{multline}\label{zeta1}
%\tag{SM3}
\zeta_n^0\left\{\left(K\frac{(nA-1)}{n^2(n-1)}^{nA-1}\right)\exp\left[(nA-1)\left(\frac{1}{K}\frac{n(n^2-1)}{nA-1}-1\right)\right]\right\}\times\\
\left\{A(\frac{n(n^2-1)}{K(nA-1)}-1)+ \frac{1}{K}\left[\frac{(nA-1)(3n^2-1)-nA(n^2-1)}{(nA-1)} \right]\right.+  \\ +\left.n\left(\frac{A}{n}-\frac{2(nA-1)}{(n^2-1)}-\frac{(nA-1)}{n^2}\right)+A\ln\left[\frac{K(nA-1)}{n(n^2-1)}\right]\right\}=0
\end{multline}
The first factor in curly brackets is strictly positive for any $n \ge 1$. Hence, Eq.~(\ref{zeta1}) is satisfied only if the second factor in curly brackets vanishes, which, in the asymptotic limit $n>>1$, yields
\begin{equation}\label{zeta2}
%\tag{SM4}
\frac{3n^2}{K}-3A+A\ln\left(\frac{KA}{n^2}\right)=0\,.
\end{equation}
Equation~(\ref{zeta2}) has two real solutions: $n_{\mathrm{A1}}=\sqrt{KA}$ and $n_{\mathrm{A2}}=\sqrt{-\frac{1}{3}W\left(\frac{-3}{e^3}\right)KA}$, where $W(x)$ is the Lambert function satisfying $x=W(x)e^{W(x)}$. Note that $n_{\mathrm{A1}} > n_\mathrm{f}$ while $n_{\mathrm{A2}} < n_\mathrm{f}$, where
\begin{equation} \label{eq:nf}
n_\mathrm{f}=\sqrt{\frac{1}{3}KA}
\end{equation}
is the mode with the maximum growth rate as obtained in Ref.~\cite{Miranda1998}. A numerical analysis of the problem shows that the number of fingers grows with time, as confirmed by the experiments. Thus, at any time the mode with maximum growth rate must be larger than that with maximum amplitude. It follows that the first solution, $n=n_{\mathrm{A1}}$, is non-physical. The final expression for the mode with the maximum amplitude is then
\begin{equation}\label{nA}
%\tag{SM5}
n_\mathrm{A}=\alpha n_\mathrm{f} \text{\hspace{1cm}          with    } \alpha=\sqrt{-W\left(\frac{-3}{e^3}\right)}\,,
\end{equation}
which is Eq.~(2) of the main text.

Equation~(\ref{nA}) has been derived in the limit $n>>1$. To test how stringent this condition actually is, we calculate numerically the exact solution to Eq.~(\ref{der0}) and compare it to the asymptotic result, Eq.~(\ref{nA}), for fluid parameters close to those of our experiments. Figure S1 shows that the agreement is indeed very good already for $n_\mathrm{A} \geqslant 2$.

\subsection{Flow curves}
The flow curves of the microgel suspensions are obtained by performing steady rate rheology experiments, using a cone-plate geometry (cone diameter = 50 mm, cone angle = 0.0198 rad) for low microgel concentrations ($Phi_\mathrm{C}\le 0.4$), and a 25 mm-plate with a roughened surface for suspensions at volume fraction $0.4<Phi_\mathrm{C}\le 1.2$, to avoid wall slip. The flow curve has been measured both by increasing sequentially the shear rate and by decreasing it, starting from its largest value. No difference are observed depending on the chosen protocol. The flow curves of the the microgel suspensions at different colloidal volume fractions $\PhiC$ (see Fig. 3 of the main text) are fitted by using functional forms issued from standard rheological models. Below, we report the functional form and the fitting parameters for all curves.
For suspensions in the jammed state we use the Herschel-Bulkley equation~\cite{Cloitre2003}:
\begin{equation}\label{HB}
\sigma(\dot{\gamma})=\sigma_y+\lambda\dot{ \gamma}^{\beta}
\end{equation}
where $\sigma(\dot{\gamma})$ is the shear stress and $\dot{\gamma}$ the shear rate. The fitting parameters are the yield stress $\sigma_y$ and the model parameters $\lambda$ and $\beta$. For $\PhiC=1.2$ and $\PhiC=0.92$ we find the following parameters:
\begin{center}
\begin{tabular}{l|l|l}
\multicolumn{1}{c|}{\textbf{$\PhiC$}} &
\multicolumn{1}{c|}{\textbf{$0.92$}} &
\multicolumn{1}{c}{\textbf{$1.2$}} \\
\hline
$\sigma_y \mathrm{(Pa)}$ & 2.50 & 8.62\\
$\lambda$ ($\mathrm{Pa~s}^\beta$)& 2.89 & 3.24\\
$\beta$ & 0.48 & 0.48\\
\end{tabular}
\end{center}

For intermediate concentrations, right below the jamming transition, the shear stress $\sigma(\dot{\gamma})$ is well described by a linear combination of two power laws:
\begin{equation}\label{DPL}
\sigma(\dot{\gamma})=A\dot{\gamma}^a+B\dot{\gamma}^b
\end{equation}
where the fitting parameters are $A$, $B$, and the exponents $a$ and $b$. The same functional form has been used to describe the flow of other soft particles, \textit{i.e.} glassy star polymer solutions in good solvent conditions~\cite{Erwin2010}. For $\PhiC=0.61$ and $\PhiC=0.55$ the values of the parameters are:
\begin{center}
\begin{tabular}{l|l|l}
\multicolumn{1}{c|}{\textbf{$\PhiC$}} &
\multicolumn{1}{c|}{\textbf{$0.55$}} &
\multicolumn{1}{c}{\textbf{$0.61$}} \\
\hline
$A$ ($\mathrm{Pa~s}^a$)& 0.06 & 0.47\\
$a$ & 0.05 & 0.10\\
$B$ ($\mathrm{Pa~s}^b$)& 0.20 & 0.33\\
$b$ & 0.50 & 0.49\\
\end{tabular}
\end{center}

As a general remark for all suspensions with $\PhiC \geq 0.55$, we note that at high shear rate the stress varies very nearly as the square root of $\dot{\gamma}$. This behavior has been proven to be universal for concentrated suspensions of soft particles~\cite{seth11}.

For $\PhiC=0.4$ the flow curve has been fitted using a Cross-like equation:
\begin{equation}\label{Cross}
\sigma(\dot{\gamma})=\frac{\eta_0 \dot{\gamma}}{1+(C\dot{\gamma})^m}
\end{equation}
where $\eta_0$ is the zero-shear viscosity of the solution, $1/C$ is the characteristic shear rate denoting the onset of the shear thinning and $m$ the shear thinning exponent. We obtain the following values for the fitting parameters:
\begin{center}
\begin{tabular}{l|l}
\multicolumn{1}{c|}{\textbf{$\Phi_C$}} &
\multicolumn{1}{c}{\textbf{$0.4$}} \\
\hline
$\eta_0$ ($\mathrm{Pa~s}$) & 0.077\\
$C$ ($\mathrm{s}$) & 0.183\\
$m$ & 0.57\\
\end{tabular}
\end{center}

Finally, for $\PhiC<0.4$ the microgel suspensions exhibit a Newtonian behavior and the flow curves are fitted via a simple linear function
\begin{equation}\label{newtonian}
\sigma(\dot{\gamma})=\eta\dot{\gamma}\,.
\end{equation}
We find:
\begin{center}
\begin{tabular}{l|l|l|l}
\multicolumn{1}{c|}{\textbf{$\Phi_C$}} &
\multicolumn{1}{c|}{\textbf{$0.20$}} &
\multicolumn{1}{c|}{\textbf{$0.26$}} &
\multicolumn{1}{c}{\textbf{$0.30$}} \\
\hline
$\eta$ ($\mathrm{Pa~s}$) & 0.0017 & 0.0023 & 0.0035\\
\end{tabular}
\end{center}

\subsection{Calculation of the square gradient (Korteweg) constant}

Balsara and Nauman~\cite{Balsara1988} derived the square gradient (or Korteweg) constant $\kappa$ introduced in Eqs. (1) and (8) of the main text from the entropy of mixing for a spatially inhomogeneous solution of Gaussian polymer chains. As we shall review it below, the calculation of Ref.~\cite{Balsara1988} crucially relies on the scaling $\ell^2 \sim a^2 N$, where $\ell$ is the typical size (\textit{e.g.} the end-to-end distance or the radius of gyration or the hydrodynamic radius) of a chain of $N$ monomers of size $a$. In this section, we show that the entropic contribution to $\kappa$ does not change, with respect to the result by Balsara and Nauman, for polymers with a different topology, provided that the same scaling $\ell^2 \sim N$ still holds. Before discussing the behavior of $\kappa$, we argue that indeed the scaling $\ell^2 \sim N$ does apply to our microgel particles. Theoretical work by Ohno~\cite{Ohno2002} supports this hypothesis: results of the renormalization-group ($\epsilon-4D$) expansion (with $D$ the spatial dimensionality) show that if a flexible polymer network made of cross-linked long chains is dissolved in a good (or theta) solvent, its squared radius of gyration scales with $N$ in the same way as a single coil, irrespective of the network structure. Experimentally, we measure by dynamic light scattering the hydrodynamic radius, $R_\mathrm{h}$, of PNIPAM microgel particles of different sizes and plot $R_\mathrm{h}^2$ as a function of $N$ in Fig. S2. The number $N$ of monomers per microgel is determined from the number density of particles, $\frac{\PhiC}{V_\mathrm{p}}$ ($V_\mathrm{p} = \frac{4}{3}\pi R_\mathrm{h}^3$ is the hydrodynamic volume of the particles), the molar mass $m_\mathrm{w}$ of a monomer and the monomer mass concentration $c$ (in units of mass per volume): $N$ = $c N_\mathrm{A}V_\mathrm{p} / (\PhiC m_\mathrm{w})$, with $N_\mathrm{A}$ Avogadro's number. The data support a power law scaling $R_\mathrm{h}^2 \sim N^\beta$, with $\beta = 1.08 \pm 0.03$, very close to the linear scaling $R^2 \sim N$.

\begin{figure}[htbp]
\centering
\includegraphics[width=0.5\columnwidth,clip]{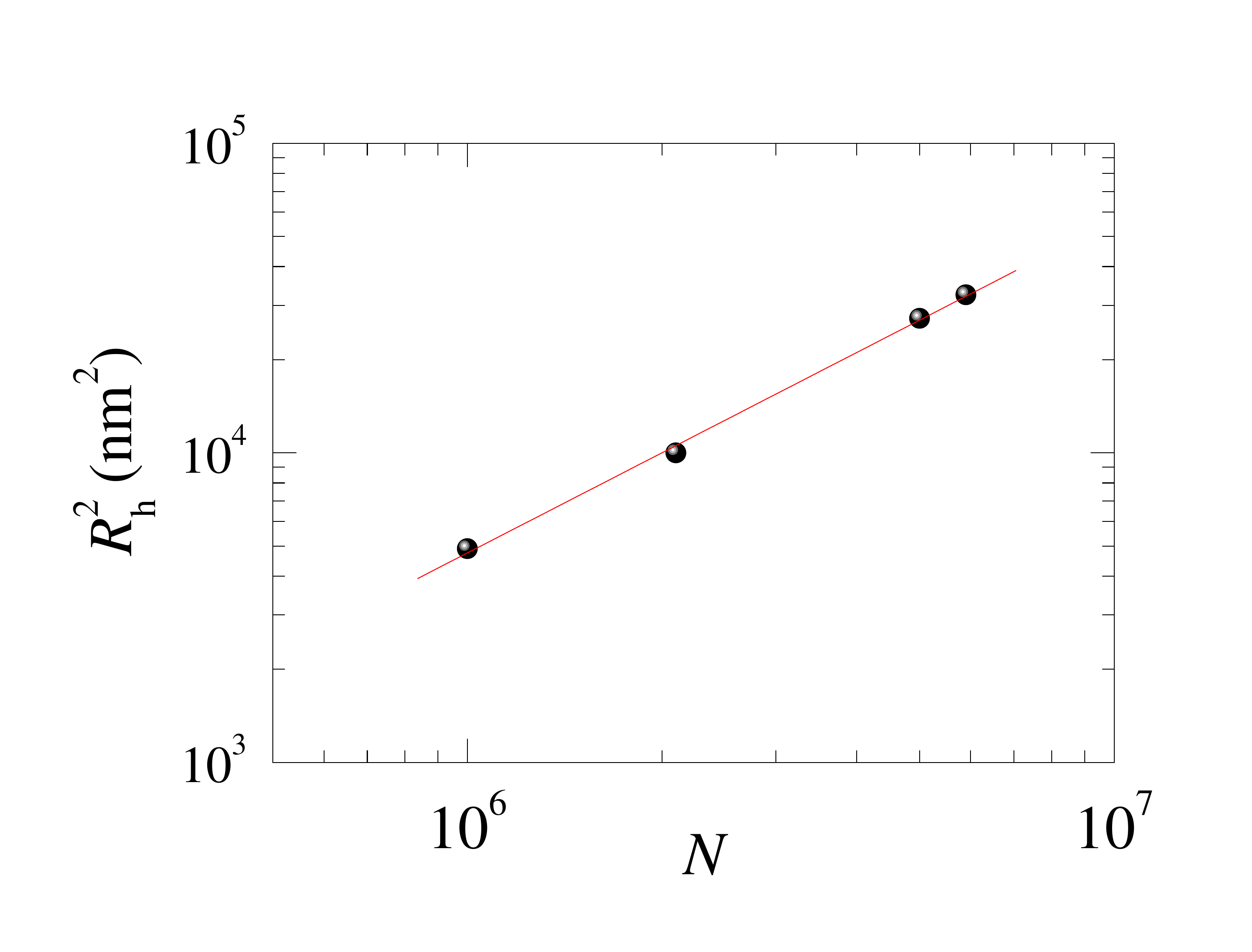}\\
\textbf{Supplementary Figure S2:} Squared hydrodynamic radius as a function of the number of monomers $N$ per microgel particle for PNIPAM microgels. The line is a power law fit, $R_\mathrm{h}^2 \sim N^\beta$, yielding $\beta = 1.08 \pm 0.03$, very close to the linear scaling $R^2 \sim N$ predicted for ideal polymers~\cite{Ohno2002}.
\label{RvsN}
\end{figure}

In order to understand how the calculation of Balsara and Nauman is modified for a cross-linked polymer, it is useful to start by recalling its main steps. $m$ chains, each of which contains $N$ monomers, occupy  a volume $V$, which is discretized in $n_0$ lattice sites; a lattice site can be occupied by one monomer at most. The local polymer volume fraction $\varphi$ at a distance $\mathbf{r}$ from a generic lattice site $P$ can be expressed as a Taylor series in terms of the volume fraction at point $P$ as follows:
\begin{equation}
\varphi(\mathbf{r})=\varphi_P+\left[(\mathbf{r}\cdot \nabla)\varphi\right]_P+\frac{1}{2}\left[(\mathbf{r}\cdot \nabla)^2\varphi\right]_P.
\end{equation}
The average volume fraction at a distance $L$ from a randomly chosen lattice site is then
\begin{equation}
\bar{\varphi}(L)=\bar{\varphi}+\frac{1}{6}(\nabla^2\varphi)_P L^2\,,
\end{equation}
where $\bar{\varphi}$ is the volume fraction averaged over all the $n_0$ sites. In order to calculate the entropy of mixing of an inhomogeneous polymer solution, one has to take into account the contribution of the concentration gradient, which modifies the probability of occupancy of a given site. Assuming that $i$ chains already occupy the lattice, for small concentration gradients the fraction of sites available to the $n$-th monomer of the $(i+1)$-th chain is given by~\cite{Balsara1988}
\begin{equation}
\label{eq:fni}
f_{ni}=1-\frac{iN}{n_0}+\frac{1}{6}(\nabla^2\varphi)l_{n}^2\,,
\end{equation}
where $l_n^2 = na^2$ is the average squared extension of a segment with $n$ monomers, with $a$ the monomer-monomer distance.
For a lattice with coordination number $z$, the number of ways of arranging the $(i+1)$-th chain, $\nu_{i+1}$, is:
\begin{equation}\label{ni-ideal}
\nu_{i+1}=(n_0-iN)\times z f_{1i} \times (z-1) f_{2i}.....\times (z-1)(f_{Ni}) \,,
\end{equation}
corresponding to a total number $\Omega_{12}$ of distinguishable arrangements of $m$ chains given by
\begin{equation}\label{Omega_12}
\Omega_{12}=\frac{1}{m!}\prod_{i=0}^{m-1}\nu_{i+1} \,.
\end{equation}
The total entropy of mixing is obtained from
\begin{equation}\label{DeltaS}
\Delta S=k_B \ln\left[\frac{\Omega_{12}}{\Omega_{1}\Omega_{2}}\right] \,
\end{equation}
where  $k_B$ is Boltzmann's constant and $\Omega_{1}$ and $\Omega_{2}$ are respectively the number of distinguishable arrangements of the polymer chains and of the solvent molecules before mixing.

We now replace the Gaussian chains of Ref.~\cite{Balsara1988} by the cross-linked chains of our microgels, adopting  a minimal model of a long Gaussian chain composed of $N$ monomers $G$ times cross-linked. In the presence of cross-links, Eqs.~(\ref{eq:fni})-(\ref{DeltaS}) need to be modified. If the scaling $\ell^2 \sim N$ holds, as for our microgels, Eq.~(\ref{eq:fni}) is still valid, provided that the average squared end-to-end distance of a Gaussian segment with $n$ monomers is replaced by the appropriate expression for cross-linked chains:
\begin{equation}
\label{eq:lnxlinked}
\widetilde{l}_n^2=ba^2 n\,,
\end{equation}
where here and in the following a tilde sign is used for variables referring to the case of cross-linked chains. In Eq.~(\ref{eq:lnxlinked}), $b$ is a suitable prefactor, whose value for the limiting case of a Gaussian coil is $b=1$. In Eq.~(\ref{ni-ideal}), the prefactors $z$ or $z-1$ account for the number of sites available to the next monomer to be placed on the lattice. For the cross-linked monomers, these prefactors will be reduced, since two monomers must be placed simultaneously on the lattice, which increases the constraints on the number of possible ways of placing them. With no loss of generality, we may assume that the number of ways of arranging the $(i+1)$-th chain is given by a modified expression,
\begin{equation}\label{ni-crosslinked}
\widetilde{\nu}_{i+1}=\Upsilon(z,b)\nu_{i+1}\,,
\end{equation}
where the prefactor $\Upsilon$ depends only on the network topology, via $b$, and the lattice coordination number. The number of distinguishable arrangements of $m$ cross-linked chains is then
\begin{equation}\label{Omega_12xlink}
\widetilde{\Omega}_{12}=\frac{\Upsilon(z,b)^m}{m!}\prod_{i=0}^{m-1}\nu_{i+1} \,,
\end{equation}
yielding the following modified expression for the total entropy of mixing
\begin{equation}\label{DeltaSxlink}
\Delta \widetilde{S}=k_B \ln\left[\frac{\widetilde{\Omega}_{12}}{\widetilde{\Omega}_{1}\Omega_{2}}\right] \,.
\end{equation}

We note that the entropy of mixing for cross-linked chains has the same formal expression as that for Gaussian chains, Eq.~(\ref{DeltaS}), the only difference being constant prefactors in $\widetilde{\Omega}_{12}$ and $\widetilde{\Omega}_{1}$, and, more crucially, $l_n^2$ being replaced by $\widetilde{l}_n^2$ in $\widetilde{\Omega}_{12}$. We thus follow the same procedure as in Ref.~\cite{Balsara1988} in order to isolate the terms in the r.h.s. of Eq.~(\ref{DeltaSxlink}) that depend on the concentration gradient. One finds that the entropy of mixing per site, $\Delta \widetilde{s} \equiv  \Delta \widetilde{S}/n_0$, reads
\begin{equation}\label{entropy1}
\Delta \widetilde{s} =k_B\left\{\frac{1}{n_0}\ln\left[\frac{\Omega_{0}}{\widetilde{\Omega}_1\Omega_2}\right]+\frac{1}{n_0}\ln\left[1-\frac{n_0 Nba^2\nabla^2\varphi}{12}\ln\left(1-\varphi\right)\right]\right\} \,
\end{equation}
with
\begin{equation}
\nonumber
\Omega_{0} = \frac{1}{m} \left [ \frac{\Upsilon(z,b)z(z-1)^{N-2}}{n_0^{N-1}} \right ] ^m \,.
\end{equation}
Equation~(\ref{entropy1}) can be further approximated by neglecting terms of order $(\nabla^2\varphi)^2$, yielding
\begin{equation}\label{entropy2}
\Delta \widetilde{s} =\Delta s_0(\varphi)-\frac{k_BNba^2}{12}\ln(1-\varphi)\nabla^2\varphi \,,
\end{equation}
where $\Delta s_0(\varphi) = \frac{k_B}{n_0}\ln\left[\frac{\Omega_{0}}{\widetilde{\Omega}_1\Omega_2}\right]$ is the entropy of mixing per site for a homogeneous solution of cross-linked polymers.

The total entropy of mixing is obtained by integrating Eq.~(\ref{entropy2}) over the volume $V$. By applying the divergence theorem and choosing a boundary $S$ such that $\int_S \nabla \varphi \cdot \mathbf{n}\mathrm{d}S = 0$ , one finds (see Eq. (2.5) in Ref.~\cite{cahn58})
\begin{equation}\label{entropy3}
\Delta \widetilde{S}=\int_V\left[\Delta s_0(\varphi)-\frac{k_BNba^2\varrho}{12(1-\varphi)}(\nabla\varphi)^2\right] \mathrm{d}V \,,
\end{equation}
where $\varrho$ is the number of sites per unit volume.
Equation~(\ref{entropy3}) can be combined with the enthalpy of mixing of an inhomogeneous solution of polymers~\cite{McMaster1975} to yield an expression for the Gibbs free energy of mixing, $\Delta G$, in the Landau-Ginzburg form
\begin{equation}\label{LandauGinzburg}
\Delta G= \int_V\left[\Delta g(\varphi)+\kappa(\nabla\varphi)^2\right] dV
\end{equation}
where $\Delta g(\varphi)$ is the density of Gibbs free energy of mixing for a homogeneous polymer solution~\cite{greiner95} and $\kappa$ is the square gradient or Korteweg constant, whose explicit expression is
\begin{equation}\label{kappa}
\kappa=\frac{RT}{V_w}\frac{Nba^2}{12}\left[\frac{\chi}{3}+\frac{1}{1-\varphi}\right]=\frac{RT}{V_w}\frac{R_\mathrm{e}^2}{12}\left[\frac{\chi}{3}+\frac{1}{1-\varphi}\right] \,,
\end{equation}
where $R$ is the gas constant, $V_w$ the molar volume of the solvent, $\chi$ the Flory-Huggins interaction parameter for the cross-linked polymer, and $R_\mathrm{e}$ the end-to-end radius of the polymer. Note that once expressed as a function of $R_\mathrm{e}$, the Korteweg constant, r.h.s. of Eq.~(\ref{kappa}), has the same expression as in Ref.~\cite{Balsara1988}.

For the sake of comparison with the experiments, it is convenient to write the Korteweg constant as a function of $R_\mathrm{g}$, the radius of gyration of the polymer. For cross-linked polymers~\cite{Solf1996},
\begin{equation}\label{gyr}
R_\mathrm{g}^2=\frac{R_\mathrm{e}^2}{6}\,,
\end{equation}
Using Eqs.~(\ref{kappa}) and (\ref{gyr}), one finds
\begin{equation}\label{kappa2}
\kappa=\frac{RT}{V_w}\frac{R_\mathrm{g}^2}{6}\left[\chi+\frac{3}{1-\varphi}\right] \,,
\end{equation}
which, in the limit $\varphi <<1 $ relevant for our experiments, coincides with Eq. (8) of the main text.

As discussed in the main text, for our microgels at temperature below the lower critical solution temperature $\chi \leq 0.5$, so that the dominant contribution to the Korteweg constant is given by the entropic term associated with the internal degrees of freedom of the polymer chains.

%\bibliographystyle{apsrev}
%\bibliography{manuscript}

\begin{thebibliography}{43}
\bibitem[{not({\natexlab{b}})}]{noteinjection}
\bibinfo{note}{In the worst case, $\Phi_\mathrm{C} = 0.2$, the longest
  injection time is $0.3~\mathrm{s}$, to be compared to $t_\mathrm{c} =
  0.6~\mathrm{s}$, the time it takes a particle to diffuse over its own
  diameter.}
\bibitem[{not({\natexlab{c}})}]{noteGamma}
\bibinfo{note}{Note that for our microgels with $R_{\mathrm h} = 160$ nm,
  $k_\mathrm{B} T/a^{-2} = 1.6 \times 10^{-4}~\mathrm{mN~m}^{-1}$, four orders
  of magnitude smaller than the typical value of $\Gamma_e$}.
\end{thebibliography}

\end{document}